**Giant negative thermal expansion covering room temperature in nanocrystalline GaN$_x$Mn$_3$**

Jianchao Lin [a], Peng Tong [a,*], Xiaojuan Zhou [b], He Lin [b,*], Yanwei Ding [c], Yuxia Bai [c], Li Chen [a], Xinge Guo [a], Cheng Yang [a], Bin Song [a], Ying Wu [a], Shuai Lin [a], Wenhai Song [a], and Yuping Sun [d,a,e,*]

[a] *Key Laboratory of Materials Physics, Institute of Solid State Physics, Chinese Academy of Sciences, Hefei 230031, China*

[b] *Shanghai Institute of Applied Physics, Chinese Academy of Sciences, Shanghai 200120, China*

[c] *Hefei National Laboratory for Physical Sciences at Microscale, University of Science and Technology of China, Hefei 230026, China*

[d] *High Magnetic Field Laboratory, Chinese Academy of Sciences, Hefei 230031, China*

[e] *Collaborative Innovation Center of Advanced Microstructures, Nanjing University, Nanjing 210093, China*

**Abstract**

Materials with negative thermal expansion (NTE), which contract upon heating, are of great interest both technically and fundamentally. Here, we report giant NTE covering room temperature in mechanically milled antiperovksite GaN$_x$Mn$_3$ compounds. The micrograin GaN$_x$Mn$_3$ exhibits a large volume contraction at the antiferromagnetic (AFM) to paramagnetic (PM) (AFM-PM) transition within a temperature window ($\Delta T$) of only a few kelvins. The grain size reduces to ~ 30 *nm* after slight milling, while $\Delta T$ is broadened to 50K. The corresponding coefficient of linear thermal expansion ($\alpha$) reaches ~ -70 ppm/K, which is almost two times larger than those obtained in chemically doped antiperovskite compounds. Further reducing grain size to ~ 10 *nm*, ΔT exceeds 100 K and *α* remains as large as -30 ppm/K (-21 ppm/K) for *x* = 1.0 (*x* = 0.9). Excess atomic displacements together with the reduced structural coherence, revealed by high-energy X-ray pair distribution functions, are suggested to delay the AFM-PM transition. By controlling the grain size via mechanically alloying or grinding, giant NTE may also be achievable in other materials with large lattice contraction due to electronic or magnetic phase transitions.

*Keywords:* Negative thermal expansion; Size effect; Nanocrystalline metals; High-energy X-ray diffraction; Magnetostructural transformations

*Corresponding Authors

*Email addresses:* tongpeng@issp.ac.cn (P. Tong), linhe@sinap.ac.cn (H. Lin), ypsun@issp.ac.cn (Y. P. Sun)

1. **Introduction**

It is well known that most materials expand upon heating and shrink upon cooling (positive thermal expansion, PTE), which consequently lead to stress concentration and mechanical fatigue in some high-precision devices (such as optical fiber reflective grating devices, printed circuit boards, and machinery parts)[1, 2]. So, the material showing the opposite thermal expansion property (i.e. negative thermal expansion, NTE), which can be mixed with PTE materials to form composites with precisely tailored coefficient of thermal expansion, receives a great deal of attention [1, 2]. Large coefficient of thermal expansion which can effectively suppress the thermal expansion of PTE matrix but with little influence on its physical properties is thus desirable for NTE material.

NTE materials studied thus far can be grouped into two main categories according to the NTE mechanisms. One type is associated with soft-phonon mechanism as a result of unusual lattice vibrations. $ZrW_2O_8$ and $ScF_3$, showing NTE up to 1000K with a coefficient of linear thermal expansion ($α$) of around -10 ppm/K, are good examples of this type [3, 4]. The other type of NTE materials involve a transition between two distinct electronic configurations or magnetic states [5, 6]. Recently, colossal NTE with $α$ lager than -70 ppm/K was reported in $Bi_{1-x}Ln_xNiO_3$ (Ln = La, Nd, Eu, Dy) [7, 8]. By partially replacing Ni with Fe, the value of $α$ can be increased to -187 ppm/K [9]. Here, the charge transformation accounts for the NTE. Another charge-transfer compound $SrCu_3Fe_4O_{12}$ shows NTE with $α$ of -22.6 ppm/K between 170 K and 270 K estimated from X-ray diffraction data [10]. High pressure of a few GPa is required to make $BiNiO_3$- and $SrCu_3Fe_4O_{12}$-based compounds, which hinders their wide-range applications [7].

Sharp magnetovolume effect (MVE) in itinerant magnetic materials, which means a sudden and large volume contraction upon heating, provides an alternative for exploring metallic NTE materials [6]. By properly doping in La(Fe,Si)$_3$-based compounds, large $α$ value of -26.1 ppm/K was achieved between 240 K and 350 K in $LaFe_{10.5}CoSi_{1.5}$ [11]. Recently, large $α$ of -51.5 ppm/K between 122 K and 332 K was obtained in anisotropic $MnCo_{0.98}Cr_{0.02}Ge$ bonded with 3-4% epoxy [12]. The residual stress originates from the bonding effect was suggested to relax the sharp MVE. More extensively MVE-related NTE has been studied for antiperovskite manganese nitrides $ANMn_3$ (where A represents the main group elements or some 3$d$ elements) [6, 13]. In parent $ANMn_3$ compounds, large MVE usually occurs at antiferromagnetic (AFM) to paramagnetic (PM) (AFM-PM) transition [6, 14]. In 2005, Takenaka et al firstly reported that the sharp MVE can be broadened in chemically doped $Cu_{1-x}Ge_xNMn_3$, giving rising to a large NTE ($\alpha$ = -16 ppm/K at 267K-342K for $x$=0.47) [15]. This observation quickly attracted research interest [16-21] because of the advantages of $ANMn_3$ compounds, such as structural isotropy, good thermal conductivity, low-cost of raw materials, and mechanical hardness [13, 15, 22]. Despite a lot of efforts made to improve NTE by manipulating the chemical compositions, achievement of both large $\alpha$ exceeding -30 ppm/K and wide temperature region over 50 K was still rare [21, 23, 24].

Here, we report giant room-temperature NTE in nanocrystalline $GaN_xMn_3$ powders prepared by ball milling (BM). The abrupt lattice contraction at the AFM-PM transition, i.e., MVE, in the non-milled samples was relaxed after BM. NTE with $\alpha$ of about -70 ppm/K and temperature interval of $\Delta T$ ~ 50 K was observed near room temperature for the samples with grain size of ~ 30 *nm*. When the average grain size diminishes to ~10 *nm*, $\Delta T$ is extended to be larger than 100 K and $\alpha$ is still considerably large (e.g., -30 ppm/K for $x$ = 1.0). The concurrent magnetic and structural transition corresponding to MVE or NTE is characterized by the coexistence of two cubic phases with distinct lattice volumes. The local structure disorder revealed by pair distribution function (PDF) is proposed to be responsible for the broadened NTE.

## 2. Experiment

Polycrystalline samples of $GaNMn_3$ and $GaN_{0.9}Mn_3$ were prepared via a direct solid-state reaction [25]. The as-prepared samples were crushed into powders and then sealed with zirconia balls and alcohol (as wet medium) in a stainless steel vial in Ar atmosphere. The ball-to-powder-to-alcohol weight ratio was 5:1:0.6. The BM process was carried out using a high-energy planetary ball mill (QM-IF) and subjected to different milling times at a constant speed (200 rpm). The as-prepared $GaN_{0.9}Mn_3$ and $GaNMn_3$ samples were denoted as N09-BM0 and N10-BM0, respectively. For ball-milled samples, the BM hours were used to name the samples. For example, N10-BM35 represents for the $GaNMn_3$ sample subjected to 35 hours of BM. Thermal behavior was investigated using a Differential Scanning Calorimeter (DSC) (TA instruments model Q2000) from 223 K to 473 K. Magnetization measurements were performed on a Quantum Design (QD) superconducting quantum interference device (SQUID) magnetometer (1.8 K ≤ $T$ ≤ 400 K, 0 ≤ $H$ ≤ 50 kOe). The surface morphology was determined using a field-emission scanning electronic microscope (FE-SEM, FEI-designed Sirion 200, Hillsboro, OR). Temperature dependent X-ray powder diffractions (XRD) between 110 K and 450 K were recorded using a Philips X′pert PRO X-ray diffractometer with Cu Kα radiation ($\lambda$ = 1.54Å). High-energy synchrotron X-ray scattering was carried out between 198K and 373K on BL13W1 in Shanghai Synchrotron Radiation Facility. X-ray energy is 69.525 keV ($\lambda$ = 0.1783Å).

### 3. Results and discussion

*3.1 Sample characterizations*

*3.1.1 X-ray diffractions* and *SEM measurements*

For $GaN_xMn_3$, the $\Gamma^{5g}$-type AFM to PM transition at $T_N$ is accompanied with a sharp change of lattice constant without change in the crystal symmetry (*Pm-3m*) [14]. During this magnetic transition, the X-ray diffraction peaks (e.g., (111) peak) of N10-BM0 (Fig.

1) and N09-BM0 (Fig. S1 in the Supplementary Material) split into two sets, indicating the coexistence of two cubic phases. The clear-cut diffraction peaks allow us to fit the diffraction patterns using two cubic phases. The refined lattice constant ($a_0$) and phase fraction were plotted in Fig. 2a-d. It is obvious that the overall lattice expands upon cooling because the large-lattice phase (ph1) grows at the expense of the small-lattice one (ph2). After BM, the diffraction peaks become wider (Fig. 1 and Fig. S1). However, the phase coexistence around $T_N$ still presents in the ball-milled samples, as manifested by the enhanced Full Width at Half Maximum (FWHM) of (111) diffraction peak (see Fig. 2e,f). Moreover, the temperature range in which the (111) peak shifts to higher angles upon heating are boradened as the BM time increases.

The average grain size ($<D>$) was estimated using the (111) X-ray diffraction peak by means of the well-known Scherrer formula [26]: $<D> = 0.93\lambda/(B\cos\theta)$, where $\lambda$ is the wavelength of the X-ray radiation, $\theta$ the diffraction angle position of the (111) peak, and $B$ is the full width at half-maximum of the peak after taking into account the instrumental peak broadening. As shown in Fig. S2 in the Supplementary Material, as the milling time increases, the grain size reduces from hundreds of *nm* for the as-prepared samples to ~ 30 *nm* for N09-BM10 and N10-BM10, and further to ~10 *nm* for N09-BM35 and N10-BM35. In agreement with the reduction of grain size, the particles broke down during BM process (Fig. 3). For example, the average particle size was diminished from 3-4 *μm* for N10-BM10 to ~1 *μm* for N10-BM35. Moreover, as milling time increases, the particles tend to be more uniform in size.

*3.1.2 DSC and magnetization measurements*

Fig. 4a shows the DSC curves measured for GaN$_{0.9}$Mn$_3$ series samples at a heating rate of 2.5K/min. The DSC peak corresponding to the magnetic/structural transition is quite sharp in N09-BM0, but broadened after BM. In addition, a small but visible shift of DSC peak to lower temperatures was observed as the milling time increases, as shown in the inset of Fig. 4a. As shown in Fig. 4b and 4c for N09-BM10 and N09-BM35, respectively, the DSC peak measured after 10 thermal cycles overlaps with the original one. The different DSC peak values recorded upon heating and cooling at 10 K/min indicates a thermal hysteresis, which suggests the structural transition under study is first order in nature. This is consistent with the fact that two cubic phases are involved in the transition (Fig. 2).

The temperature dependent magnetization $M$(T) for the GaNMn$_3$ series samples were measured at 100 Oe under zero-field-cooling warming, field-cooling cooling and field-cooling warming modes. The AFM-PM transition is clearly visible on both $M$(T) and 1/$M$(T) curves for the non-milled sample (N10-BM0), as shown in Fig. 5a and 5b. After BM, however, the transition becomes less distinguishable. Eventually, the signals on $M$(T) and 1/$M$(T) curves due to AFM-PM transition is no longer noticeable for N10-BM35 sample. Fig. 5c displays the field dependent magnetization, $M$(H), measured at 5 K for all N10 series samples. The non-milled sample (N10-BM0) exhibits a linear field dependence of magnetization, consistent with the AFM ground state. With increasing BM time, a weak FM component was introduced as evidenced by a slightly increased magnetization at 45 kOe and an increase of magnetic coercivity. Such a weak ferromagnetism can be attributed to the uncompensated surface spins, which has been often observed in nanocrystalline AFM materials, such as MnO [27].

*3.2 Giant negative thermal expansion*

Based on the refined lattice constant (Fig. 2a-b), the linear thermal expansion $\Delta a_0/a_0$(321 K) was calculated for the non-milled samples and displayed in Fig. 6a,e. Upon heating, sharp volume contraction occurs at ~315 K and 347 K with $\Delta T$ ~8 K for N10-BM0 and N09-BM0, respectively. The lattice change at $T_N$ for N10-BM0 is about $4.1 \times 10^{-3}$, which is close to the reported value (~$4.3 \times 10^{-3}$) measured using dilatometric method [6].

For the ball-milled samples, the double X-ray diffraction peaks belong to the two phases are hard to be distinguished (Fig. 1 and Fig. S1) and thus treated as a single peak during structure refinement. Based on the refinement results, the calculated $\Delta a_0/a_0$(321 K) was plotted in Fig. 6b-h. As compared with the non-milled sample, $\Delta T$ is increased to 53 K (271 K-324 K), 80 K (243 K-323 K), and 103 K (221 K-324 K) for N10-BM10, N10-BM25 and N10-BM35 samples, respectively. Accordingly, the average $\alpha$ is -76.1 ppm/K, -43.6 ppm/K, -30.1 ppm/K for N10-BM10, N10-BM25 and N10-BM35, respectively. The N09 series samples exhibit a similar behavior, i.e., the temperature window for NTE becomes wider and the NTE coefficient decreases gradually with increasing BM time. The N09-BM10, N09-BM22, N09-BM35 samples show an average $\alpha$ of -68.7 ppm/K, -46.8 ppm/K, -21.5 ppm/K at 305 K - 357 K ($\Delta T$ = 52 K), 292 K - 360 K ($\Delta T$ = 63 K), and 250 K - 360 K ($\Delta T$ = 110 K), respectively.

The $\alpha$ values of slightly ball-milled $GaN_xMn_3$ (N10-BM10 and N09-BM10) are almost two times larger than those reported in chemically doped $ANMn_3$ [17, 21, 23, 24]. Moreover, such large $\alpha$ values are comparable to that observed in $Bi_{0.95}La_{0.05}NiO_3$ (-82 ppm/K) [7], but much larger than that observed in $LaFe_{11.5}Si_{1.5}$ (-50 ppm/K) [11]. It is noteworthy that the two-phase coexistence survived in our BM samples is absent in the chemically doped ANMn$_3$ showing NTE. The coexisting two phases with different lattice volumes have also been suggested to be crucial for the giant NTE observed in Bi$_{0.95}$La$_{0.05}$NiO$_3$ [7]. For heavily milled samples (e.g., N10-BM35, N09-BM35), the NTE temperature range can be larger than 100 K with $\alpha$ larger than -20 ppm/K. The reversibility of NTE as demonstrated by the DSC thermal cycling experiment (Fig. 4b, c) is vital for practical applications. Small particle size (~ 1 $\mu m$) of ball-milled samples is beneficial to composite with materials whose positive thermal expansion needs to be reduced under control [2]. In view of those advantages discussed above, the current nanocrystalline antiperovskite manganese nitrides may have potential applications as a thermal-expansion compensator.

*3.3. High-energy X-ray Pair distribution functions, G(r)s*

Atomic PDF, derived from high-energy X-ray or neutron scattering data which considers both the Bragg and diffuse scattering intensities, provides information on short-range atom displacements that are not necessarily periodic in real space [28]. The PDF analysis has been widely used as a local structural probe to explore the microstructure related to NTE mechanisms [1]. The high-energy X-ray scattering data were measured for N10-BM0 and N10-BM25 to address the change of local structure due to BM. The data were subjected to routine corrections and background subtraction, and finally the atomic PDF, $G(r)$s, were obtained through a Fourier transform [28]. All data were processed under identical conditions using the same termination $Q_{max}$ (22 Å$^{-1}$) as well as the same background correction.

*3.3.1. G(r) peak broadening beyond thermal vibrations*

All the *G*(*r*)s were plotted in Fig. S3 in the Supplementary Material. For the purpose of clarity, in Fig. 7 we only plotted the *G*(*r*)s at 198 K and 373 K for both compounds. Splitting of the *G*(*r*) peak around 2.75 Å observed in chemically doped antiperovskite manganese nitrides is invisible here. Instead, the PDF peaks for the ball-milled sample are suppressed in intensity and shifted towards lower *r* values, both of which are particularly noticeable in the higher *r* region. The reduced peak intensity which accumulates with distance indicates that the local structure loses its coherence on a larger scale [29, 30]. To take a deeper insight into the change of PDF after BM, the peaks at ~2.75 Å, 4.76 Å and 7.30 Å were fitted using a Gaussian profile. The extracted peak position and peak width were plotted in Fig. 8 for all *G*(r)s. The peak position (i.e., bond length) behaves as the average lattice does. For each sample, the peak width increases with temperature because of the increasing thermal vibrations. For N10-BM0, the peak widths exhibit an abnormal at $T_N$ which agrees with the abrupt volume change. While for N10-BM25, the peak widths (especially at 4.76 Å and 7.30 Å) show a broad hump in the NTE temperature range, which can be attributed to the coexisting two cubic phases as revealed by the X-ray diffraction data (Fig. 2). Furthermore, the PDF peaks for N10-BM25 are substantially wider than those for N10-BM0 in the whole temperature range studied. Thermal effect can be safely excluded since such extra PDF peak broadening is temperature independent.

*3.3.2. Modeling of local structure*

In order to understand the underlying physics of the PDF peak broadening after BM, we tested the *G*(*r*)s at 198 K and 373 K for both N10-BM0 and N10-BM25 samples using the average cubic model (space group, *Pm-3m*). It turns out that the cubic model fits all the data well (see Fig. 9a,b for the fitting at 373 K). For *Pm-3m* symmetry, the thermal factor (atomic displacements) is isotropic at both the Ga and N sites ($U_{11} = U_{22} = U_{33}$), but anisotropic at the Mn site [28]. However, the refined $U_{11}$ ($U_{22}$) at Mn sites is very close to $U_{33}$, no matter whether the sample was milled or not. The rather isotropic atomic displacements again exclude the local symmetry breaking [28]. The obtained atomic displacements (Table S1 in Supplementary Material) at Mn and Ga sites are enhanced after BM, while those for N sites are decreased. The atomic displacements for N site may be less accurate than those for Mn and Ga sites because the X-ray atomic scattering factor for N is much smaller than that of Mn and Ga. So, the broadening in PDF peaks caused by BM originates from static displacements of atoms rather than local symmetry breaking or thermal vibration. The enhanced atomic disorder beyond thermal vibrations and reduced structural coherence observed in N10-BM25 can be readily attributed to the diminished grain size [29].

It was observed previously that, in nanocystalline $Cu_{0.5}Ge_{0.5}NMn_3$ [18], $Cu_{0.6}Ge_{0.4}NMn_3$ [31] and $Zn_{0.6}Ge_{0.4}NMn_3$ [24] compounds, a reduction of grain size to nanoscale leads to a further broadening of NTE temperature span. Accordingly, the value of *α* was dramatically reduced, even to ~ 0.1 ppm/K (i.e., the so-called zero thermal expansion) [18]. This is not surprising because the magnitude of volume change is quite small ($< 2 \times 10^{-3}$) in the related coarse-grain or micro-grain samples which already have a well broadened NTE temperature range (> 30 K) [18, 24, 31]. In the above reports, the issues of local structure have not been addressed. Instead, a reduction of Mn occupancy from 100% in the coarse-grained sample to 78.7% was suggested to modulate the NTE behavior of ultra-nanocrystalline $Cu_{0.5}Ge_{0.5}NMn_3$ [32]. Meanwhile, a downward shift as large as ~ 110 K was found for the NTE temperature range [18, 32], which is much larger than the values observed in our samples and in nanocrystalline $Cu_{0.6}Ge_{0.4}NMn_3$ and $Zn_{0.6}Ge_{0.4}NMn_3$. If Mn deficiencies do exist, the nearest Mn-Mn bond should split into three parts, two longer ones and a shorter one [32]. However, the PDF peaks at 2.7 Å and 3.85 Å arising from Mn-Mn and Mn-Ga bonds show a clear Gaussian shape (Fig. 9c for N10-BM25 at 373K), indicating no Mn-Mn bond splitting. Furthermore, the *Pm-3m* cubic model with Mn vacancies fails to fit the experimental $G(r)$ (the case of 10% Mn vacancies is shown in Fig. 9d as an example).

*3.4. Relation between NTE and static atom displacements*

For chemically doped $ANMn_3$, the broadening of MVE was believed to stem from the slow-down or the disturbance of the growth of the $\Gamma^{5g}$ AFM order [6]. A three-dimensional web consists of corner-shared $Mn_6N$ octahedral forms in the antiperovskite structure. When the nearest Mn-Mn interaction is AFM-type, spin frustration is formed [33]. The long-range $\Gamma^{5g}$ AFM order can only be stabilized in a very narrow energy window with the assistance of the strong ferromagnetic (FM) next-nearest-neighbor interaction [33]. So, the $\Gamma^{5g}$ AFM order can be easily disturbed [25]. The random displacements of Mn away from their equilibrium positions revealed by our PDF data suggest a dispersed distribution of Mn-Mn bond lengths. The AFM exchange integrals are thus spread out in strength, which would result in a broadened AFM-PM transition [28, 34, 35]. Furthermore, the lattice tends to lose coherence after BM, which prevents the propagation of the $\Gamma^{5g}$ AFM order when cooled from high temperatures. As a combined effect, the growth of the $\Gamma^{5g}$ AFM order would be impeded, leading to the broadening of the magnetic transition and thus to NTE. Besides, the presence of high-degree structural disorder usually reduces the strength of the magnetic interactions, and thus shifts magnetic transition towards lower temperatures [34, 36]. It is indeed the case. For example, the mean value of $T_N$ (determined as middle point of the NTE temperature span on the $\Delta a_0/a_0$(T) curve) is reduced from 347 K for N09-BM0 to 333 K for N09-BM35 (Fig. 6). The decrease of $T_N$ is also manifested by the shift of DCS peak (Fig. 4a). The structural disorder observed here is quite universal for materials with reduced grain size [29, 30, 37]. Therefore, high-energy mechanical milling or alloying, which can reduce sample's grain size for certain, may open a new avenue for achieving giant NTE from the giant lattice contraction that is electronically or magnetically driven.

## 4. Conclusion

In summary, giant NTE covering room temperature was observed in mechanically milled $GaN_xMn_3$ powders. By reducing the grain size to nanoscale, either giant NTE with $\alpha$ ~-70 ppm/K and $\Delta T$ ~50 K, or large NTE with $\alpha$ ~-30 ppm/K and $\Delta T$ ~100 K were obtained. The giant NTE is characterized by the coexistence of two cubic phases which have distinguishable lattice volumes. The atomic displacements beyond usual thermal vibrations along with the reduced structural coherence were proposed to be responsible for the broadening of MVE. The current study suggests that in compounds of sharp lattice shrinkage concomitant with electronic or magnetic transitions, giant NTE could be obtained by mechanically grinding or alloying.

**Acknowledgements**

This work was supported by the National Key Basic Research under Contract No. 2011CBA00111; the National Natural Science Foundation of China (NSFC) under

Contract Nos. 51322105, 11174295, 51301167, 51171177, 51371005 and 91222109. HL acknowledges the NSFC under contract No. U1232112 and help from high-energy synchrotron X-ray scattering measurements on BL13W1 in Shanghai Synchrotron Radiation Facility.

**Figure captions:**

Fig. 1. The (111) peak in XRD patterns at different temperatures for the samples including (a) non-milled (N10-BM0), (b) 10-hours milled (N10-BM10), (c) 25-hours milled (N10-BM25), (d) 35-hours milled (N10-BM35) samples. The data corresponding to the temperature range of negative thermal expansion were plotted in red for each sample. The dotted lines trace the shift of peak as temperature changes.

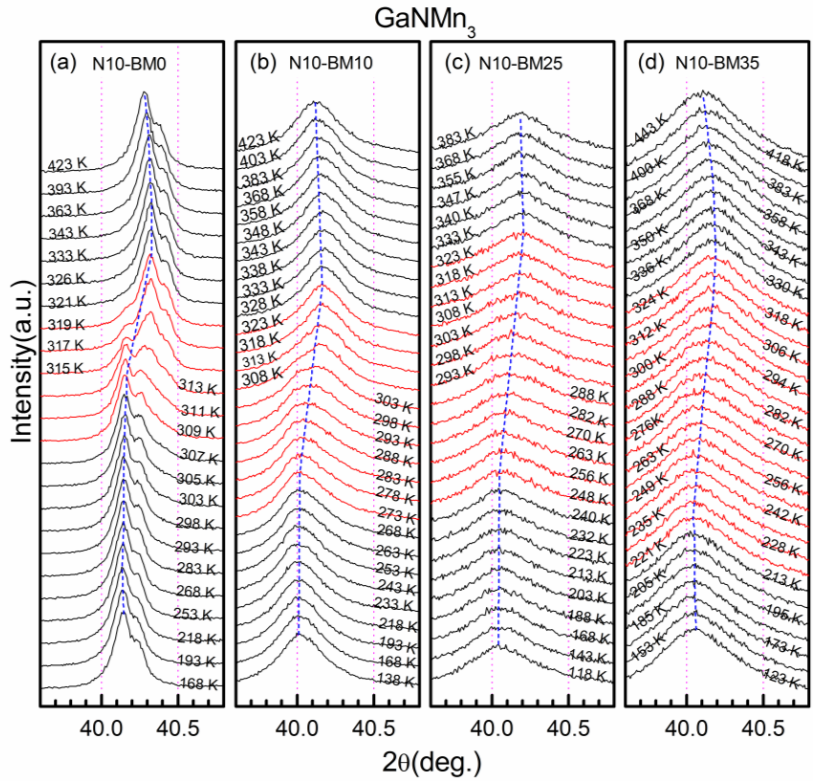

Fig. 2. Coexistence of two cubic phases at the antiferromagnetic to paramagnetic transition in $GaN_{0.9}Mn_3$ (N09) and $GaNMn_3$ (N10) series samples. For non-milled samples (N09-BM0 and N10-BM0), the refined lattice constant and phase fraction are plotted in (a)-(d). Full Width at Half Maximum (FWHM) of (111) X-ray diffraction was plotted in (e) and (f) for milled N09 (N09-BM10, N09-BM22 and N09-BM35) and N10 (N10-BM10, N10-BM25 and N10-BM35) samples, respectively.

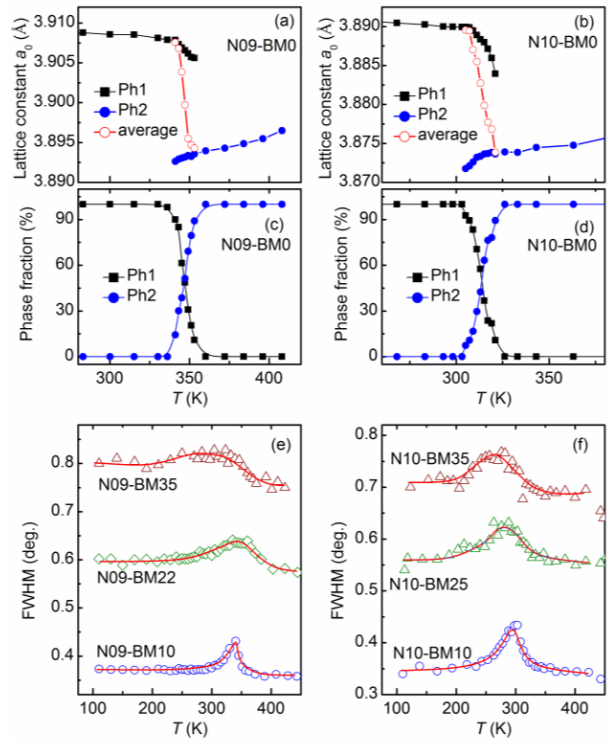

Fig. 3. The FE-SEM images for the GaNMn$_3$ (N10) samples subjected to ball-milling for 10 hours (N10-BM10) (a), 25 hours (N10-BM25) (b), and 35 hours (N10-BM35) (c).

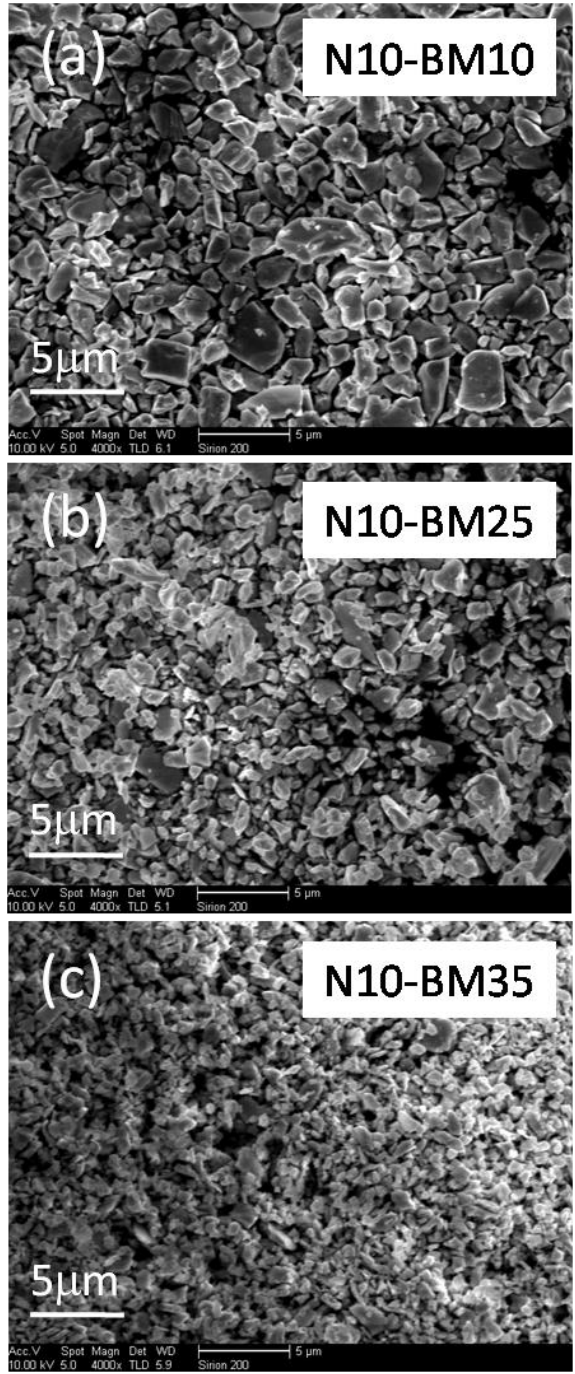

Fig. 4. The temperature dependent heat flow (DSC) between 223 K and 473 K for GaNMn$_3$ (N09) samples. (a) DSC curves measured at a heating rate of 2.5 K/min for non-milled (N09-BM0) sample and ball-milled samples with milling time varying from 4.5 hour (N09-BM4.5) to 35 hours (N09-BM35). Inset: the magnified pattern between 305K and 390K. (b) and (c) are the DSC curves recorded during 10 thermal cycles for N09-BM10 and N09-BM35, respectively. The heating or cooling rate is 10 K/min.

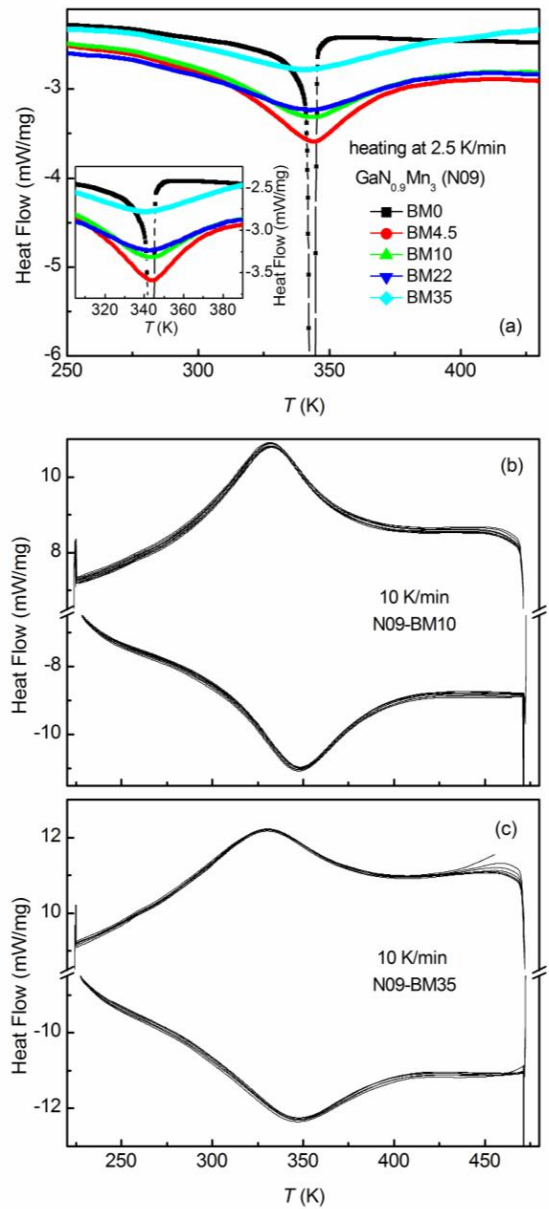

Fig. 5. The magnetization for GaNMn$_3$ (N10) series samples subjected to ball milling (BM) of 0 hour (BM0) to 35 hours (BM35). (a) The temperature dependent magnetization *M*(T) measured at 100 Oe under zero-field-cooling (ZFC) warming (red), field-cooling (FC) cooling (blue) and FC warming (black) modes. (b) The inverse susceptibility of ZFC-*M*(T), 1/*M*(T). The arrows indicate the antiferromagnetic to paramagnetic transition. (c) Field scanning of magnetization at 5 K.

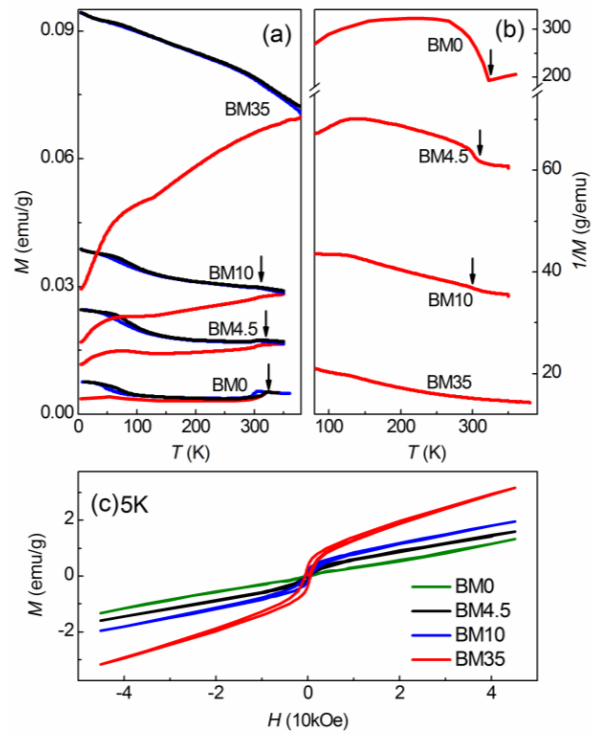

Fig. 6. Temperature dependent linear thermal expansions $\Delta a_0/a_0(321K)$ (the reference temperature is 321K) for $GaNMn_3$ and $GaN_{0.9}NMn_3$ subjected to different time periods of ball milling. The average coefficient of linear thermal expansion, $\alpha$, and the relative temperature range, $\Delta T$, were listed in each panel.

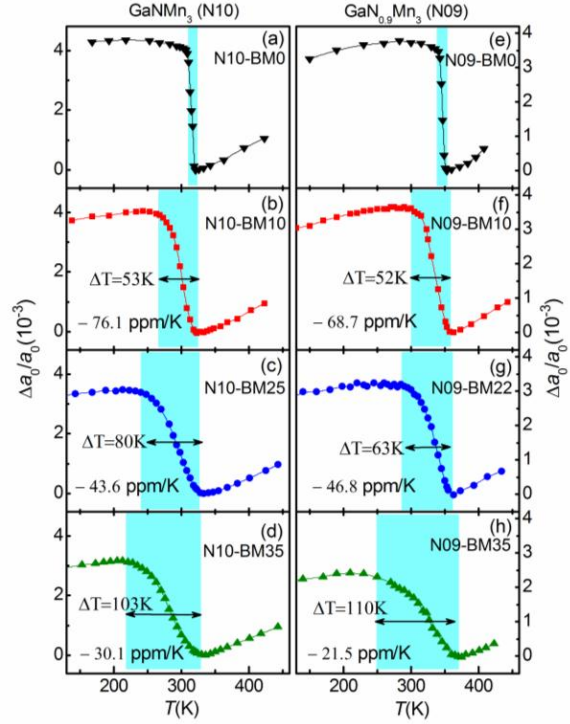

Fig. 7. $G(r)$s at 198K (a) and 373K (b) for GaNMn$_3$ before (N10-BM0) and after 25-hours (N10-BM25) of ball milling. The difference of $G(r)$s before and after milling is shown at the bottom in each panel.

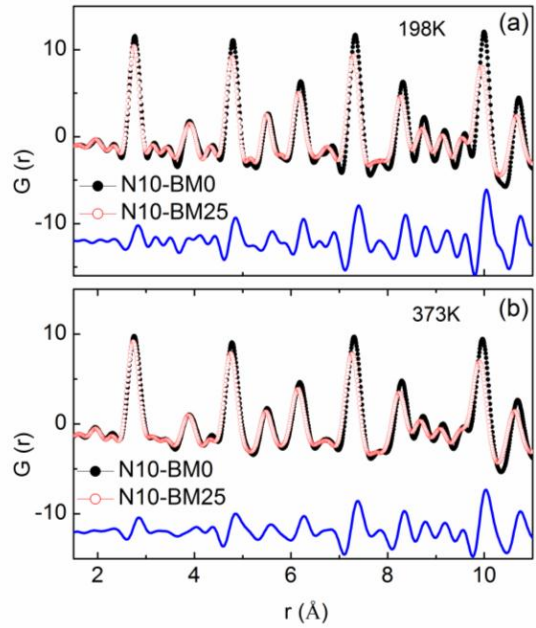

Fig. 8. Peak position (a-c) and peak width (d-f) at ~ 2.75Å, 4.76 Å and 7.30 Å in $G(r)$ at various temperatures for the non-milled (N10-BM0) and 25-hours ball-milled (N10-BM25) GaNMn$_3$ samples. See Fig. S3 in Supplementary Material for $G(r)$s data.

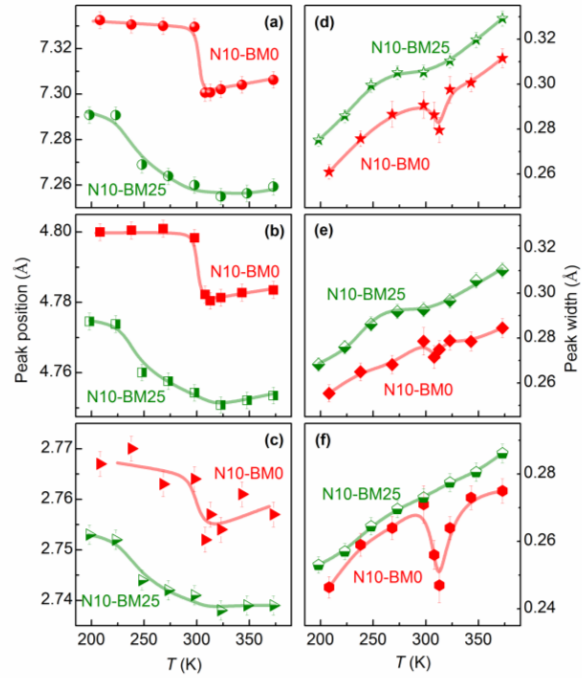

Fig. 9. Modeling of $G(r)$. Calculated $G(r)$s using cubic *Pm-3m* model (the red line) together with the experimental $G(r)$ (circle) at 373K for non-milled (N10-BM0) (a) and 25-hours milled (N10-BM25) (b) GaNMn$_3$ samples. (c) A Gaussian fit (the red line) to the experimental $G(r)$ (i.e., the data of (b)) peaks at 2.7 Å and 3.85 Å. (d) Calculated $G(r)$ using cubic *Pm-3m* model assuming 10% Mn vacancies (red line) and the experimental $G(r)$ (circle). The difference between calculated and experimental $G(r)$s is shown at the bottoms of (a), (b) and (d).

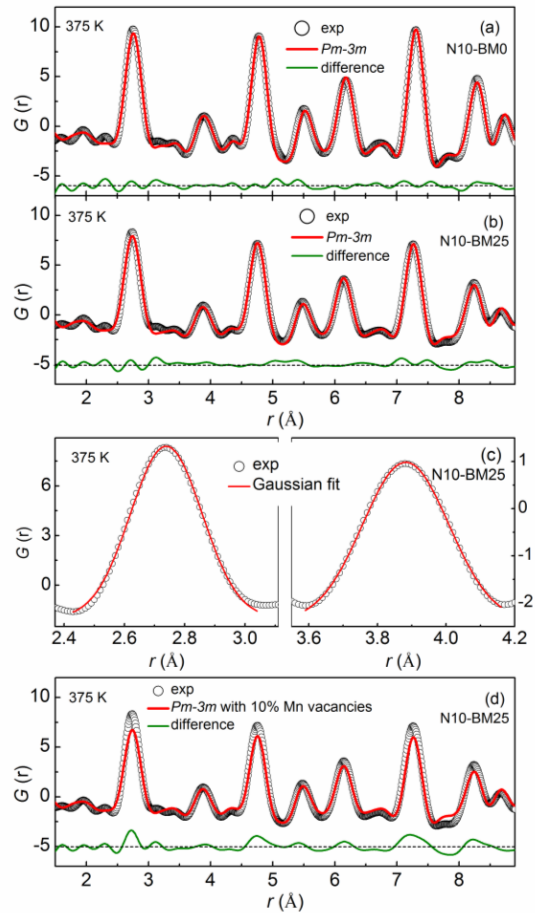

# Supplementary material

**Giant negative thermal expansion covering room temperature in nanocrystalline GaN$_x$Mn$_3$**

Jianchao Lin [a], Peng Tong [a,*], Xiaojuan Zhou [b], He Lin [b,*], Yanwei Ding [c], Yuxia Bai [c], Li Chen [a], Xinge Guo [a], Cheng Yang [a], Bin Song [a], Ying Wu [a], Shuai Lin [a], Wenhai Song [a], and Yuping Sun [d,a,e,*]

[a] *Key Laboratory of Materials Physics, Institute of Solid State Physics, Chinese Academy of Sciences, Hefei 230031, China*

[b] *Shanghai Institute of Applied Physics, Chinese Academy of Sciences, Shanghai 200120, China*

[c] *Hefei National Laboratory for Physical Sciences at Microscale, University of Science and Technology of China, Hefei 230026, China*

[d] *High Magnetic Field Laboratory, Chinese Academy of Sciences, Hefei 230031, China*

[e] *Collaborative Innovation Center of Advanced Microstructures, Nanjing University, Nanjing 210093, China*

*Corresponding Authors

*Email addresses:* tongpeng@issp.ac.cn (P. Tong), linhe@sinap.ac.cn (H. Lin), ypsun@issp.ac.cn (Y. P. Sun)

**Fig. S1.** The (111) peak in XRD patterns at different temperatures for samples including (**a**) non-milled (N09-BM0), (**b**) 10-hours milled (N09-BM10), (**c**) 22-hours milled (N09-BM22), (**d**) 35-hours milled (N09-BM35) samples, respectively. The data corresponding to the temperature range of negative thermal expansion were plotted in red for each sample. The dotted lines trace the shift of peak as temperature changes.

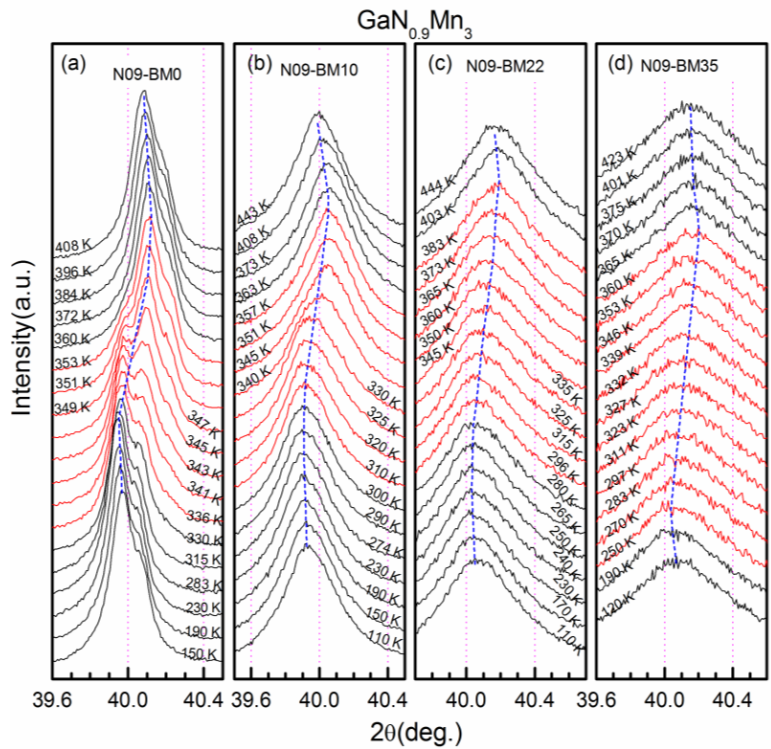

**Fig. S2.** The average grain size deduced from the (111) X-ray diffraction peak as a function of milling time for the samples GaNMn$_3$ (N10) and GaN$_{0.9}$Mn$_3$ (N09).

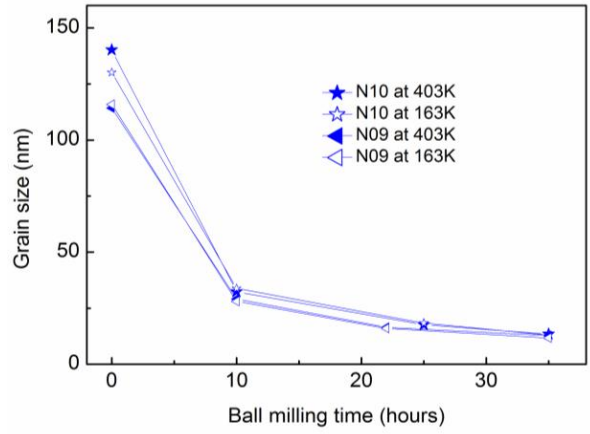

**Fig. S3.** Pair distribution functions, $G(r)$s, for non-milled (N10-BM0) (**a**) and 25-hours milled (N10-BM25) (**c**) GaNMn$_3$ samples at different temperatures. The magnified pattern of the peak at 7.3Å for N10-BM0 (**b**) and N10-BM25 (**d**).

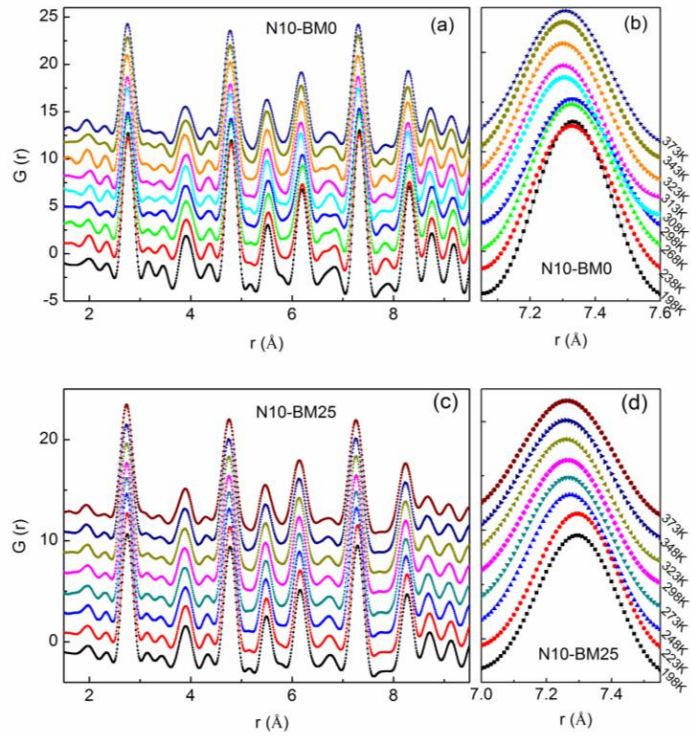

**Table S1.** The refined lattice constant, atomic displacements ($U_{11}$, $U_{22}$ and $U_{33}$) and the agreement factor ($R_w$) by fitting the $G(r)$ data for non-milled samples (N10-BM0 and N09-BM0) and 25-hours milled samples (N10-BM25 and N09-BM25) with the cubic model (space group, *Pm-3m*).

|  | 373K | | 198K | |
|---|---|---|---|---|
|  | **N10-BM0** | **N10-BM25** | **N10-BM0** | **N10-BM25** |
| **Lattice constant(Å)** | 3.911 | 3.887 | 3.922 | 3.902 |
| **Ga($U_{11}$,$U_{22}$,$U_{33}$) (Å$^2$)** | 0.0060 | 0.00675 | 0.00314 | 0.0045 |
| **N($U_{11}$,$U_{22}$,$U_{33}$) (Å$^2$)** | 0.0099 | 0.0079 | 0.01 | 0.0067 |
| **Mn($U_{11}$, $U_{22}$) (Å$^2$)** | 0.0080 | 0.0091 | 0.00588 | 0.00622 |
| **Mn($U_{33}$) (Å$^2$)** | 0.0081 | 0.0089 | 0.00592 | 0.00625 |
| **$R_w$** | 8.8% | 10.4% | 8.3% | 9.8% |